\documentclass{appolb}
\usepackage{graphicx}
\usepackage[utf8]{inputenc}
\usepackage{amssymb,amsmath}
\usepackage{authblk}
\usepackage[left=0.7in, right=0.7in, top=1in, bottom=1in]{geometry}
\usepackage{hyperref}

\begin{document}
\title{Thermodynamics of ideal Fermi gas under generic power law potential in $d$-dimension
}
\author{Mir Mehedi Faruk$^{a}$, G. M. Bhuiyan$^{b}$
\address{Department of Theoretical Physics, University of Dhaka, Bangladesh}\\
\href{mailto:me@somewhere.com}{Email: muturza3.1416@gmail.com$^{a}$, mehedi.faruk.mir@cern.ch$^{a}$, gbhuiyan@du.ac.bd$^{b}$} 
\\
}
\maketitle

 \begin{abstract}
Thermodynamics of ideal Fermi gas trapped in an external generic power law potential 
$U=\sum_{i=1} ^d c_i |\frac{x_i}{a_i}|^{n_i}$
are investigated systematically from the grand thermodynamic potential in $d$ dimensional space. 
These properties are explored deeply in the degenerate limit ($\mu>> K_BT$),  where the thermodynamic properties
are greatly dominated by Pauli exclusion principle.  Pressure and energy along with 
the isothermal compressibility
is non zero at $T=0K$, denoting trapped 
Fermi system is quite live even at absolute zero temperature.
The nonzero value of compressibilty denotes zero point pressure is not a constant but depends on 
volume.
\end{abstract}
\PACS{03.75.Ss, 05.30.-d, 05.30.Fk}
  
\section{Introduction}
The constrained role of external potential can change the characteristic of quantum 
gases\cite{turza,sala,jellal,li,dal,toroidal}.
An increasing attraction towards this subject was noticed, 
after it was possible to create
Bose-Einstein condensation (BEC) in magnetically trapped alkali gases\cite{Bradley,anderson,davis}. 
A lot of studies are performed to understand 
the behaviour of ideal Bose gas\cite{ziff,pathria,huang,C. J. PETHICK,acharyya} as well as ideal 
Fermi gas\cite{pathria,huang}. 
But unlike the Bose gas, Fermi gas does not condensate
 due to the Pauli exclusion
principle, the question of a large number of particles occupying a single energy state does
not even arise in this case.
At sufficiently low temperatures, Fermi gas displays its own brand
of interesting behavior\cite{pathria,huang}, as 
its fugacity $z_F$ can take on unrestricted values: $0 < z_F < \infty$\cite{pathria} unlike the Bose gas who has 
restricted value of fugacity $0 < z_B \leqslant 1$\cite{pathria}. 
The behavior of thermodynamic quantities of Fermi gas are remarkably governed by Pauli exclusion principle.
For instance, the ground state pressure also known as degeneracy pressure\cite{pathria} in ideal Fermi gas 
is nonzero unlike Bose gas or classical gas\cite{pathria}. Nevertheless, a lot of efforts are made to understand
different properties of  Fermi system such as magnetism\cite{acharyya2}, conductivity\cite{acharyya3},
transport properties\cite{italy},
equivalence with ideal Bose gas \cite{M. Howard Lee, pathria1}, 
dimensionality effects\cite{M. Apostol}, degeneracy\cite{demarco},
polylogarithmis\cite{lee2}, q-deforemed syetm\cite{q,q1}.
\\\\
In a real systems, of 
course interaction between particles does exist. But taking it into account makes the problem difficult to
solve analytically. Neveretheless, to understand the effect
of interactions in quantum gases and to retain
the essential physics, we approximately represent the real system by non interacting particles
in the presence of an external potential\cite{sala,thomas,bhuiyan}.
The trapping potential in atomic gases provide the opportunity to manipulate the quantum statistical effects. Some
drastic changes are noted in case of Bose system\cite{turza,sala} under trapping potential.
For instance BEC is possible in $d=2$ in trapped Bose gas which was not a case in ideal Bose gas\cite{turza,sala}.
Therefore, it will be interesting to investigate how the trapping potential do change the properties of Fermi gas.
Li et al.\cite{Li}, in their work has presented 
internal energy, heat capacity, ground state energy of Fermi gas under spherically 
symmetric potential ($U=br^t$) in arbitrary dimension.
However, in the present study we have investigated 
properties of ideal Fermi gas under generic power law potential $U=\sum_{i=1} ^d c_i |\frac{x_i}{a_i}|^{n_i}$
in $d$-dimension, which will be symmetric under certain  choice $n_i$, $x_i$ and $c_i$. So, we can recounstruct the results of 
Li et al. \cite{Li} choosing this special condition.
At first, we have calculated the density of states which enables us to determine the grand potential.
Then from the grand potential of the system, 
we have derived the thermodynamic quantities such as 
internal energy $E$
entropy $S$, pressure $P$ , number of particle $N$,
helmholtz free energy $A$, isothermal compressibility $\kappa_T$, specific heat at 
constant volume $C_V$ and pressure $C_P$ and  their ratio.
In the high temperature limit, the thermodynamic quantities of free quantum gases
reduce to the form of classical gas\cite{pathria}. 
Same trend is  observed in case of trapped system\cite{turza} too.
Therefore, the low temperature limit of the thermodynamic quantities  of quantum gases
are particularly important,  as the 
true quantum nature is explicit in that region.
In Bose system, low temperature limit refers to condensed phase. It
was found trapping potential changes the general criterion of BEC as well as the condition of
jump of specific specific heat\cite{turza,sala}. 
So, low temperature limit of thermodynamic quantities related to trapped Fermi system have 
been investigated using Sommerfield expansion\cite{pathria}.
It will be very intriguing 
to investigate  energy and pressure of trapped Fermi gas 
while the gas is in
the degenerate 
limit  to check whether they remain nonzero under generic trapping potential at $T=0K$.
Isothermal compressibility (inverse of bulk modulus) is also calculated to check whether the ground state 
pressure has volume dependency or it is marely just a constant. 
A point to note that, in the hamiltonian
instead of $\frac{p^2}{2m}$ type kinetic part, 
we have took $ap^s$, where $p$ is momentum, 
$a$ is constant and $s$ is a arbitrary kinematic parameter. Different kinematic characteristics
of quantum systems lead to different characteristics\cite{turza,beckmann,beckmann2}. 
So, choosing arbitrary kinematic parameter, we make
the significant
conclusions in more generalized way. In the current study it is found
the concept of effective volume plays an important role in trapped Fermi gas, 
as seen in trapped Bose gas\cite{turza}. \\\\
The report is organized in the following way. The density of states and grand potential are calculated in section 2.
Section 3 is devoted to investigate the thermodynamic quatities. Properties of degenerate Fermi gas 
is presented in section 4.
Results and
discussions are presented in section 5. The report is concluded in section 6.

\section{Density of States and grand potential of Fermi gas under generic power law potential in d dimension}
Considering  the ideal  Fermi gas with  kinematic parameter $l$ in 
a confining external potential in a d-dimensional space with energy spectrum, 
\begin{eqnarray}
\epsilon (p,x_i)= bp^l + \sum_{i=1} ^d c_i |\frac{x_i}{a_i}|^{n_i}
\end{eqnarray}
Where, $b,$ $l,$ $a_i$, $c_i$, $n_i$  are all postive constants, $p$ is the momentum 
and $x_i$ is the  $i$ th component of coordinate of a particle. Here, $c_i$, $a_i$ and $n_i$ determines the depth 
and confinement power of
the potential. With $l=2$, $b=\frac{1}{2m}$ one can get the energy spectrum  of  the hamiltonian used in 
the literatures \cite{pathria,huang,ziff,sala}.
For the free system all $n_i\longrightarrow \infty$. \\\\
Density of states can be obtained from the following formula,
\begin{eqnarray}
 \rho(\epsilon)&=& \int \int \frac{d^d r d^d p}{(2 \pi \hslash)^d} \delta(\epsilon - \epsilon (p,r))
\end{eqnarray}\\
So, from the above equation density of states is,
\begin{eqnarray}
\rho(\epsilon)=B\frac{\Gamma(\frac{d}{l} + 1)}{\Gamma(\chi) }\epsilon^{\chi-1} 
\end{eqnarray}
where, 
\begin{eqnarray}
B=\frac{gV_d C_d}{h^d a^{d/l}}\prod_{i=1} ^d \frac{\Gamma(\frac{1}{n_i} + 1)}{c_i ^{\frac{1}{n_i}}}
 \end{eqnarray}\\
Here,  $C_d=\frac{\pi^{\frac{d}{2}}}{\Gamma(d/2 + 1)} $, $g$ is the spin degeneracy factor,
$V_d=2^d\prod_{i=1} ^d a_i$ is the volume 
of an $d$-dimensional rectangular
whose $i$-th side has length $2a_i$. 
$\Gamma(l)=\int_0 ^\infty dx x^{l-1}e^{-x}$ is the gamma function and  $\chi= \frac{d}{l} + \sum_{i=1} ^d \frac{1}{n_i}$.\\\\
The grand potential for the Fermi system, 
\begin{eqnarray}
q=-\sum_\epsilon ln(1+zexp(-\beta \epsilon)) 
\end{eqnarray}
$\beta=\frac{1}{kT}$,
where $k$ being the Boltzmann Constant and $z=\exp(\beta \mu)$ is the fugacity, where $\mu$ being the chemical potential.
Using the Thomas-Fermi semiclassical 
approximation\cite{thomas} and  re-writing the previous equation,
\begin{eqnarray}
 q=q_0 -\int_0 ^\infty \rho(\epsilon) ln(1+zexp(-\beta \epsilon))
\end{eqnarray}
So, using the density of states of Eq. (3) we finally get the grand potential,
\begin{eqnarray}
q=q_0+ B \Gamma(\frac{d}{l}+1)(kT)^\chi f_{\chi+1} (z)
\end{eqnarray}
where, $q_0=-ln(1-z)$
and  $f_l(z)$ is the Fermi function which is defined as,
\begin{equation}
 f_p(z)=\int_0 ^\infty dx\frac{x^{p-1}}{z^{-1}e^x+1}=\sum_{j=1} ^\infty (-1)^{j-1}\frac{z^p}{j^p}
\end{equation}

\section{Thermodynamics of Fermi gas under generic power law potential in d dimension}
\subsection{Number of Particle}
The number of particles $N$ can be obtained,
\begin{eqnarray}
 N=z(\frac{\partial q}{\partial z})_{\beta,V}= N_0 + \frac{gC_n\Gamma(\frac{d}{l} +1)
 V_d \prod_{i=1} ^d  \Gamma(\frac{1}{n_i}+1 ) }{h^d b^{d/l} \prod_{i=1} ^d c_i ^{1/n_i} } (kT)^\chi f_\chi(z)
\end{eqnarray}
Here, $N_0=\frac{z}{1-z}$ is the ground state occupation number. \\ \\ 
Now, defining, \begin{eqnarray}
                V_d ' &=& V_d \prod_{i=1} ^d (\frac{kT}{c_i})^{1/n_i}\Gamma(\frac{1}{n_i} + 1)\\
                \lambda'&=& \frac{h b^{\frac{1}{l} }}{\pi ^{\frac{1}{2}} (kT) ^{\frac{1}{l}}} [\frac{d/2+1}{d/l+1}]^{1/d}
               \end{eqnarray}\\\\
It is noteworthy,
\begin{eqnarray}
&&\lim_{n_i\to\infty} V_d'=V_d  \\
&&\lim_{n_i\to\infty} \chi =\frac{d}{l}\\
&&\lim_{l\to 2, b\to \frac{1}{2m}} \lambda' =\lambda_0=\frac{h}{(2\pi mk T)^{1/2}}
\end{eqnarray}\\
So, if we choose $l=2$ and $b=\frac{1}{2m}$ from Eq. (14) 
we get $\lambda_0=\frac{h}{(2\pi mk T)^{1/2}}$, which is the thermal 
wavelength of nonrelativistic free massive fermions. However, it should be noted that, 
when $l\neq 2$,  $\lambda'$ 
then depends on dimension. With $d=3$ and $d=2$, thermal wavelength of photons are respectively $\frac{hc}{2\pi^{1/2} kT}$ 
and $\frac{hc}{(2\pi)^{1/2}kT}$ which can be obtained from from Eq. (11) by choosing $b=c$, where $c$ being
the velocity of light. So, one can reproduce the thermal wavelength of  both massive and massless fermions 
from the definition of $\lambda '$ with more general energy
spectrum. But one needs to consider the effects of 
antiparticles to calculate the thermodynamic quantities of ultrarelativistic quantum gas\cite{howard}.\\ \\
The number of particle equation is then written as,
\begin{equation}
N-N_0= g \frac{V_d '}{{\lambda '} ^d}f_\chi(z)
\end{equation}
The number of particle equation for free
massive fermions (with $l=2$, $a=\frac{1}{2m}$, all $n_i\longrightarrow \infty$) in $d$ dimesnsional space can be 
obtained from Eq. (15),  
\begin{equation}
 N-N_0= g \frac{V_d}{{\lambda_0} ^d}f_\frac{d}{2}(z)
\end{equation}
which gives the exact equation for
number of particles at $d=3$\cite{pathria,huang}.\\\\
\subsection{Internal Energy}
From the Grand Canonical Ensemble
internal energy $E$  is, \begin{eqnarray}
                             E&=&-(\frac{\partial q}{\partial \beta})_{z,V} \nonumber  \\
                              &=& \frac{gC_n\Gamma(\frac{d}{l} +1) V_d \prod_{i=1} ^d  
                              \Gamma(\frac{1}{n_i}+1 ) }{h^d b^{d/l} \prod_{i=1} ^d c_i ^{1/n_i} } (kT)^{\chi+1} f_{\chi+1}(z)   \\  
     &=&        NkT \chi \frac{f_{\chi+1}(z)}{f_{\chi}(z)}      
    \end{eqnarray}
In case of free massive fermions,                            
  \begin{eqnarray}
   E    &=&  NkT \frac{d}{2} \frac{f_{d/2+1}(z)}{f_{d/2}(z)}
\end{eqnarray}                       
which is in accordance with the exact expression of $E$
for $d=3$\cite{pathria,huang}.\\\\
Now as $T\longrightarrow \infty$, from Eq. (19) it is seen,
the internal energy becomes, $E=NkT\chi$.
For free massive fermions it is $E=\frac{d}{2}NkT$, which becomes $\frac{3}{2}NkT$,
when $d=3$. Thus $E$ approaches the classical value at high temperature. The exact 
same trend is also seen in case of Bose 
gas\cite{turza}.\\
\subsection{Entropy}
The entropy $S$ can be obtained from Grand Canonical Ensemble, 
\begin{eqnarray}
 S&=&kT(\frac{\partial q}{\partial T})_{z,V} -Nk\ln z +kq\nonumber\\  
  &=&N k [\frac{v_d'}{\lambda '^d}(\chi+1) {f_{\chi + 1}(z)}-\ln z]
 \end{eqnarray}
As before, for free massive fermions, one can find Eq. (20) approaches to,
\begin{eqnarray}
   S = N k [\frac{v_d}{\lambda ^d}(\frac{d}{2}+1) {f_{\frac{d}{2} + 1}(z)}-\ln z]      
\end{eqnarray}
Again at $d=3$, Eq. (21)
reduces to same expression for entropy as Ref.\cite{pathria,huang}\\
\subsection{Helmholtz Free Energy}
From the Grand Canonical Ensemble we get the expression of Helmholtz Free Energy for Fermi system,
\begin{eqnarray}
 A&=&-kTq+NkT\ln z\nonumber\\
&=&      - NkT \frac{f_{\chi+1} (z)}{f_{\chi} (z)} +  NkT\ln z
 \end{eqnarray}
In case of free massive fermions, the above expression reduces like below,
\begin{eqnarray}
   \frac{A}{NkT}= -\frac{f_{\frac{d}{2}+1} (z)}{f_{\frac{d}{2}} (z)}+\ln z 
\end{eqnarray}         
 Now, for $d=3$, the above equation
produces the exact expression for Helmholtz free Energy\cite{pathria,huang}.
\subsection{Pressure}
Rewriting equation (15) stating the number of particles,
\begin{equation}
\frac{N-N_0}{V_d \prod_{i=1} ^d (\frac{kT}{c_i})^{1/n_i}\Gamma(\frac{1}{n_i} + 1)}=\frac{N-N_0}{V_d '}=\frac{g}{\lambda '^d}g_{\chi}(z)  \nonumber
\end{equation}\\
Now a very well known expression for the nonrelativistic $d-$dimensional ideal free Free gas\cite{pathria},
\begin{equation}
\frac{N-N_0}{V_d}=\frac{g}{\lambda_0 ^d} f_{d/2}(z) \nonumber
\end{equation}\\
Comparing the above
equations, we can say $V_d '$ is a more generalized
extension of  $V_d$. Where, 
\begin{eqnarray}
V_d ' = V_d \prod_{i=1} ^d (\frac{kT}{c_i})^{1/n_i}\Gamma(\frac{1}{n_i} + 1) \nonumber
\end{eqnarray}
It represents the effect of external potential on the performence of 
trapped fermions. Calling $V_d '$ the effective volume the grand potential can be rewritten as,
\begin{equation}
 q=q_0+g\frac{gV_d'}{\lambda '^d}f_{\chi+1}(z)
\end{equation}\\
So, the effective pressure \begin{eqnarray}
                            P'=\frac{1}{\beta}(\frac{\partial q}{\partial V_d '})=\frac{gkT}{\lambda '^d} f_{\chi+1} (z) 
                           \end{eqnarray}
Which can be rewritten as,
\begin{eqnarray}
                            P'=     \frac{NkT}{V_d '} \frac{f_{\chi+1} (z)}{f_{\chi}(z)} 
                           \end{eqnarray}\\
The above equation is very general equation of state
for any dimensionality $d$, any dispersion relation 
of the form $(\propto$ $p^s$) having
any form of generic power law trap and obviously it is expected that it will reproduce the 
special case of free system. For free system the equation (26) becomes,
\begin{eqnarray}
                            P=\frac{1}{\beta}(\frac{\partial q}{\partial V_d })
 =      \frac{NkT}{V_d } \frac{f_{d/2+1} (z)}{f_{\frac{d}{2}}(z)} 
                           \end{eqnarray}
   which is in accordance with Ref.\cite{pathria,huang} at d=3.                        
                           \\\\
                           Now, comparing Eq. (18) and (26) we get,
\begin{equation}
 P'V_d '=\frac{E}{\chi}
\end{equation}
For $d$-dimensional free Fermi gas one can obtain from previous equation
\begin{equation}
 P V_d=\frac{2}{d}E
\end{equation}
This is an important and familiar relation, $P V=\frac{2}{3}E$ when $d=3$\cite{pathria,ziff,huang,C. J. PETHICK}.
This actually shows equation (28) is a very significant relation
for the Fermi gas irrespective whether they are trapped or free.
And in case of trapped fermions effective volume and effective pressure
plays the same role as volume and pressure in current textbooks and literatures.
Interestingly, the Bose gas also maintains this equation.\cite{turza}\\\\
\subsection{Heat Capacity}
Heat capacity at constant volume $C_v$,
\begin{eqnarray}
   C_V &=& T(\frac{\partial S}{\partial T})_{N, V} \nonumber \\
        & = &  N k [\chi(\chi+1)\frac{\nu'}{\lambda '^D}f_{\chi+1}(z)-\chi^2 \frac{f_{\chi}(z)}{f_{\chi-1}(z)}] 
\end{eqnarray}         
For Free massive fermions, the expression becomes,
\begin{eqnarray}
   C_V  =   N k [\frac{d}{2}(\frac{d}{2}+1)\frac{\nu}{\lambda ^D}f_{\frac{d}{2}+1}(z)-{\frac{d}{2}}^2 \frac{f_{\frac{d}{2}}
   (z)}{f_{\frac{d}{2}-1}(z)}] 
\end{eqnarray}         
And
in the high temperature limit of $C_v$ approches its classical value as it becomes $\chi Nk$ for trapped
system and $\frac{d}{2}Nk$ for free system, which is $\frac{3}{2}Nk$, when $d=3$. \\\\
Now, heat capacity at constant pressure $C_p$, 
\begin{eqnarray}
   C_P &=&T(\frac{\partial S}{\partial T})_{N, P} \nonumber\\
      &=& Nk  [   (\chi+1)^2 f_{\chi+1}^2(z)f_{\chi-1}(z)
      (\frac{\nu'}{\lambda '^D})^3-\chi(\chi+1)f_{\chi+1}(z)\frac{\nu'}{\lambda '^D}] 
\end{eqnarray}         
In case of free massive Fermi gas, the above equation reduces to,
\begin{eqnarray}
C_P = Nk [   (\frac{d}{2}+1)^2 f_{\frac{d}{2}+1}^2(z)f_{\frac{d}{2}-1}(z)(\frac{\nu'}{\lambda '^D})^3-\frac{d}{2}(\frac{d}{2}+1)f_{\frac{d}{2}+1}(z)\frac{\nu'}{\lambda '^D}] 
\end{eqnarray}\\
It coincides exactly with Ref. \cite{pathria,C. J. PETHICK} for $d=3$. Again in the high 
temperature limit $C_p$ becomes $(\chi+1)Nk$ for trapped
system and $(\frac{d}{2}+1)Nk$ for free system, which is $\frac{5}{2}Nk$, when $d=3$. So, in 
the high temperature limit  $C_p$ approches its classical value.\\\\
Now, the ratio, $\gamma=(\frac{C_P}{C_V})$ for Fermi gas is given by,  
\begin{eqnarray}
  \gamma=\frac{(\chi+1)^2 \frac{f_{\chi+1}^2(z)f_{\chi-1}(z)}{f^3 _\chi (z)}-\chi(\chi+1)\frac{f_{\chi+1}(z)}{f_\chi(z)} } 
  {   \chi(\chi+1)\frac{\nu'}{\lambda '^D}f_{\chi+1}(z)-\chi^2 \frac{f_{\chi}(z)}{f_{\chi-1}(z)}}
\end{eqnarray}\\
Now, in high temperature limit, the above equation becomes,
\begin{eqnarray}
 \gamma=\frac{(\chi +1)^2-\chi(\chi +1)}{\chi(\chi +1)-\chi^2}=1+\frac{1}{\chi}
\end{eqnarray}
In case of free system, choosing all $n_i\longrightarrow \infty$, we get from the above equation,
\begin{equation}
\gamma=1+\frac{l}{d} 
\end{equation}
With $d=3$ and $l=2$, $\gamma$ equals $\frac{5}{3}$, 
thus obtaining the  classical value at high temperature limit.
\subsection{Isothermal Compressibility}
The Isothermal compressibility of Fermi gas can be obtained,
\begin{eqnarray}
\kappa_T&=&-V_d '(\frac{\partial V'}{\partial P'})_{_{N,T}}\nonumber \\
&=&-V_d '(\frac{\partial P'}{\partial z})_{_{N,T}}(\frac{\partial z}{\partial V'})_{_{N,T}}\nonumber \\
&=& \frac{V_d '}{NkT}\frac{f_{\chi-1}(z)}{f_\chi(z)} 
\end{eqnarray}
which reproduces the same result for isothermal compressibility of free massive Fermi gas at $d=3$ \cite{pathria}.
And as $T\longrightarrow \infty$ $\kappa_T$ takes the classical value for free system, which is $\frac{1}{P}$.\\ \\
\section{The thermodynamic properties of a degenerate Fermi gas under generic power law potential}
At low temperature, can approximate the Fermi
function and write it as quickly convergent Sommerfield series\cite{pathria}
\begin{eqnarray}
 f_p(z)=\frac{(\ln z) ^p}{\Gamma(p+1)}[1+p(p-1)\frac{\pi^2}{6}\frac{1}{(\ln z)^2}+p(p-1)(p-2)(p-3)\frac{7 \pi^4}{360}\frac{1}{(\ln z)^4}+...]
\end{eqnarray}
At $T=0K$, taking only the first  the first term in Eq (38). Substituting this into Eq. (15) we get 
\begin{equation}
 N-N_0=N_e =\frac{gC_n\Gamma(\frac{d}{l} +1)
 V_d \prod_{i=1} ^d  \Gamma(\frac{1}{n_i}+1 ) }{h^d b^{d/l} \prod_{i=1} ^d c_i ^{1/n_i} \Gamma(\chi+1)} E_{F}^{\chi} 
\end{equation}
Which turns out,
\begin{equation}
 E_F =[\frac{ h^d b^{d/l} \prod_{i=1} ^d c_i ^{1/n_i} \Gamma(\chi+1) N_e }{ gC_n\Gamma(\frac{d}{l} +1)V_d \prod_{i=1} ^d  \Gamma(\frac{1}{n_i}+1 )}]^{\frac{1}{\chi}} 
\end{equation}
Following the method of Ref.\cite{li,pathria,huang} we approximate
the chemical potential and fugacity from Eq.(15),
\begin{eqnarray}
\mu=kT\ln z=E_F[1-(\chi-1)\frac{\pi^2}{6}(\frac{kT}{E_F})^2] 
\end{eqnarray}\\
Using these approximation we can calculate the thermodynamic quantities of the previous section,
\begin{eqnarray}
&& \frac{E}{N} = \frac{\chi}{\chi +1} E_F [1+(\chi +1) \frac{\pi ^2}{6}(\frac{kT}{E_F})^2]\\
&& \frac{S}{Nk}= \frac{\chi \pi^2}{3 E_F} kT\\
&& P=\frac{E_F N}{(\chi+1)V'}[1+(\chi +1) \frac{\pi ^2}{6}(\frac{kT}{E_F})^2]\\
&&\frac{C_V}{Nk}= \frac{\chi \pi^2}{3 E_F} kT \\
&& \kappa_T = \frac{V' \chi}{NE_F}\{1+(1-\chi)\frac{\pi^2}{6}(\frac{kT}{E_F})^2\}
\end{eqnarray}
In case of free massive fermions (choosing $l=2$), Eq. (42)-(46) becomes,
\begin{eqnarray}
&& \frac{E}{N} = \frac{d}{d +2} E_F [1+(\frac{d}{2} +1) \frac{\pi ^2}{6}(\frac{kT}{E_F})^2]\\
&& \frac{S}{Nk}= \frac{d \pi^2}{6 E_F} kT\\
&& P=\frac{2E_F N}{(d+2)V'}[1+(\frac{d}{2} +1) \frac{\pi ^2}{6}(\frac{kT}{E_F})^2]\\
&&\frac{C_V}{Nk}= \frac{d \pi^2}{6 E_F} kT \\
&& \kappa_T = \frac{V' d}{2NE_F}\{1+(1-\frac{d}{2})\frac{\pi^2}{6}(\frac{kT}{E_F})^2\}
\end{eqnarray}
At temperature $T=0K$, entropy $S=0$ which is accordance will 3rd law of Thermodynamics. 
The internal energy, pressure and isothermal compressibility $T=0K$,
\begin{eqnarray}
E_0&=&\frac{\chi}{\chi+1}NE_F\\
P_0&=& \frac{1}{(\chi+1)}\frac{N}{V'}E_F\\
{\kappa_T}_0 &=& \frac{V \chi}{NE_F}
\end{eqnarray}
In case of Free massive Fermi gas the above equations reduces to,
\begin{eqnarray}
E_0&=&\frac{d}{d+2}NE_F\\
P_0&=& \frac{2}{(d+2)}\frac{N}{V}E_F\\
{\kappa_T}_0 &=& \frac{V }{NE_F}\frac{d}{2}
\end{eqnarray}\\
At $d=3$, Eq. (55) and (56) become exactly same as in Ref. \cite{pathria}
\section{Discussion}
Thermodynamics of ideal Fermi gas in the presence of an external generic power law potential  are
discussed in this section.  It is seen the effective volume $V_d '$
is a very salient feature of trapped system, playing the same role in trapped system
as the volume in free system, which enables us to treat trapped fermi gas as well 
the bose gas\cite{turza} to be treated as a free one. Difference between $V_d'$  and $V_d$
is that, the former depends on temperature and power law exponent while the latter does not.
But as all $n_i\longrightarrow \infty$, $V_d'$ approaches $V_d$.
In this process the more general thermal wavelength $\lambda '$ is defined with arbitrary
kinematic parameter in any dimension.  It was shown how  $\lambda '$ can reproduce the thermal wavelengths of literatures
in different dimensions. In case of trapped Fermi gases,
$V_d'$ and  $\lambda '$ enable
all the thermodynamic functions of the system to be expressed in a compact form
similar to those of free Fermi gas. \\\\ 
At first the density of states and grand potential is calculated in section 2. All the thermodynamic quantities
are derived from the grand potential in section 3.
It is seen that all the thermodynamic quantities for trapped Fermi
gas are some function of Fermi functions, just as in  the case of free Fermi gas.
But in the former case the Fermi functions depend on $\chi= \frac{d}{l} + \sum_{i=1} ^d \frac{1}{n_i}$ and $z$,
whether in the later case Fermi
functions depend on $\frac{d}{l}$ and $z$. And as  
$n_i \longrightarrow \infty$, the mathematical form of thermodynamic quantities of trapped system reduce to
that of free system. Nevertheless it is noteworthy that Eq. (28)
is a very remarkable relation for quantum gases as both Bose\cite{turza} and Fermi system maintains it.
\\\\
\begin{table}[h!]
\caption{Dissimilarity between free and harmonic-potential-trapped
nonrelativistic Fermi gases in $d=3$.}
\begin{tabular}{ |p{6.9cm}|p{4.5cm}|p{4.5cm}|  }
\multicolumn{3}{}{} \\
\hline
Physical quantity     & Free gas & Trapped gas  \\
\hline
Fermi Energy & $\frac{\hslash^2}{2m}(6\pi^2(\frac{N}{V})^{\frac{2}{3}})$ & $\hslash \omega (3N)^{1/3}$ \\
Fermi Temperature &  $\frac{\hslash^2}{2mk}(6\pi^2(\frac{N}{V})^{\frac{2}{3}})$  &  $\frac{\hslash \omega}{k}  (3N)^{1/3}$\\
Internal energy & $\frac{3}{2}NkT\frac{f_\frac{5}{2} (z)}{f_\frac{3}{2} (z)}$ & $3NkT\frac{f_4 (z)}{f_3 (z)}$ \\
Internal energy at lower temperature & $\frac{3}{5}NE_F(1+\frac{5}{2}\frac{\pi^2}{6}(\frac{kT}{E_F})^2)$ & $\frac{3}{4}NE_F(1+\frac{2\pi^2}{3}(\frac{kT}{E_F})^2)$ \\
Ground state energy    & $\frac{3}{5}NE_F$ & $\frac{3}{4}NE_F$ \\
Ground state pressure & $\frac{2}{5}\frac{N}{V}E_F$ & $\frac{1}{4}\frac{N}{V'}E_F$ \\
Particle number at ground state & $\frac{4\pi V}{3h^3}(2mE_F)^{3/2}$ & $\frac{1}{3}(\frac{E_F}{\hslash \omega})^3$   \\
Internal energy at higher temperature & $\frac{3}{2}NkT$ & $3NkT$ \\
Isothermal Compressibility & $\frac{V}{NkT}\frac{f_{1/2}(z)}{f_{3/2}(z)}$ & $\frac{V'}{NkT}\frac{f_{2}(z)}{f_{3}(z)}$ \\
Isothermal Compressibility at higher temperature & $\frac{1}{P}$ & $\frac{1}{P'}$ \\
Isothermal Compressibility at lower temperature & $\frac{3V}{2NE_F}[1-\frac{\pi ^2}{12}(\frac{kT}{E_F})^2]$ & $\frac{3V}{NE_F}[1-\frac{\pi ^2}{3}(\frac{kT}{E_F})^2]$ \\
Isothermal Compressibility at $T=0K$ & $\frac{3}{2}\frac{V}{NE_F}$ & $\frac{3V}{NE_F}$ \\
Chemical potential at lower temperature &       $E_F (1-\frac{\pi ^2}{12} (\frac{kT}{E_F})^2$)   & $E_F (1-\frac{\pi ^2}{3} (\frac{kT}{E_F})^2$)\\
Chemical potential at $T=0K$ &       $E_F $            & $E_F $\\
Chemical potential at higher temperature &       $-kT\ln [\frac{3}{4}\sqrt{\pi}(\frac{kT}{E_F})^{3/2}]$          & $-kT\ln [6(\frac{kT}{E_F})^{3}]$)\\
$\frac{C_V}{Nk}$ &  $(\frac{15}{4})\frac{f_{\frac{5}{2}}(z)}{f_\frac{3}{2} (z)}-(\frac{9}{4})\frac{f_{\frac{3}{2}}(z)}{f_\frac{1}{2} (z)}$    &  $12\frac{f_4{(z)}}{f_3 (z)}-9\frac{f_3(z)}{f_2 (z)}$                 \\
$\frac{C_V}{Nk}$ at lower temperature & $\frac{\pi^2}{2}\frac{kT}{E_F}$ & ${\pi^2}\frac{kT}{E_F}$\\
$\frac{C_V}{Nk}$ at higher temperature & $\frac{3}{2}$ & $3$\\
\hline
\end{tabular}
 \end{table}\\\\\\
In general, the thermodynamic quantities of 
trapped system differ from free system. We can specifically check this by comparing
free system with harmonically trapped 
potential. Let, $d=3$,
$a=\frac{1}{2m}$, $l=2$,  $n_1=n_2=n_3=2$, $c_i=\frac{m\omega ^2}{2}$ and $g=2$. 
Results of some of the physical quantities have  been listed in Table 1.
From the table it is seen,
thermodynamic quantities are affected by the trapping potential. The signature
of trapping potential is  present in the 
low as well as in the high temperature limit of thermodynamic functions.
As we know chemical potential approaches Fermi
Energy  as $T\longrightarrow 0$
in case of free Fermi gas, same phenomena is also observed in case of trapped Fermi gas,
although Fermi Energy do vary comparing
trapped system with the free one.
Let us turn our attention to low temperature limit of Fermi gas. It is seen both $C_V$
 and $S$ has same numerical value in this limit just like the free system
and goes to zero at $T=0K$.
The later actually is a manifestation third law of thermodynamics. 
 But most significantly internal energy 
and pressure of trapped Fermi gas
do not
go to zero as $T=0K$ just as free Fermi gas. According to equation (46), no 
matter what power law exponent
one chooses, ground state pressure never goes to zero. It suggests, 
the ground state energy and ground state pressure seen here is clearly
a quantum effect arising due to Pauli exclusion principle due to which the 
system can not settle down into a single energy state
as in the case of Bose gas. And therefore spread over a lowest available energy states. 
More interestingly, isothermal compressibility of Fermi system is nonzero 
at $T=0K$, indicating zero point pressure is not 
merely a constant but depends on volume. 
\section{Conclusion}
From the grand potential, the thermodynamic properties of Fermi
gas trapped under generic power law potential have been evaluated. The 
calculated physical quantities reduce
to expressions available in the literature, with appropiate
choice of power law exponents and dimensionality. 
The thermodynamic quantities
are studied further closely in the degeneracy limit.
It is found pressure, energy 
and isothermal compressibility is nonzero, with any trapping potential,
indicating the
governing power of Pauli exclusion principle.
In
this manuscript, we have restricted our discussion in case of ideal system under trapping potential.
It will be vey interesting 
to see the effect of interaction in the degeneracy limit.
\section{Acknowledgement}
I would like to thank Fatema Farjana, for her efforts to help me present this work and Mishkat Al Alvi 
for showing the typographic mistakes.

\end{document}